\documentclass[useAMS,usenatbib,a4paper]{mn2e}

\usepackage{graphics}
\usepackage{graphicx}
\usepackage{times}
\usepackage{amsmath}
\usepackage{natbib}
\usepackage{wasysym}
\usepackage[usenames]{color}
\usepackage{epsfig}
\usepackage{epstopdf}
\usepackage{subfigure}
\usepackage{graphics}
\usepackage{longtable}
\usepackage{amssymb}
\usepackage{latexsym} 
\usepackage{wrapfig}
\usepackage{mathrsfs}
\usepackage{morefloats}
\usepackage{bm}
\usepackage[T1]{fontenc}
\usepackage{aecompl}

\long\def\symbolfootnote[#1]#2{\begingroup%
\def\thefootnote{\fnsymbol{footnote}}\footnote[#1]{#2}\endgroup}

\title[Wavefront Sensing from the Image Domain]
{A Demonstration of Wavefront Sensing and Mirror Phasing from the Image Domain}
\author[B. J. S. Pope et al.]
{Benjamin Pope$^{1,2}$\thanks{E-mail: benjamin.pope@astro.ox.ac.uk}, Nick Cvetojevic$^{1,3,4}$, Anthony Cheetham$^{1}$,
Frantz Martinache$^{5}$,
\newauthor Barnaby Norris$^{1}$, Peter Tuthill$^{1}$  \\
$^{1}$Sydney Institute for Astronomy, School of Physics, University of 
Sydney, NSW 2006, Australia\\
$^{2}$Astrophysics, University of Oxford, Denys Wilkinson Building, Keble Rd, Oxford OX1 3RH, UK. \\
$^{3}$Centre for Ultrahigh bandwidth Devices for Optical Systems (CUDOS), Institute of Photonics and Optical Science (IPOS), \\ School of Physics, University of Sydney, Sydney, NSW 2006, Australia\\
$^{4}$Australian Astronomical Observatory, NSW 2121, Australia\\
$^{5}$Laboratoire Lagrange, CNRS UMR 7293, Observatoire de la C\^{o}te d'Azur, Bd de l'Observatoire, 06304 Nice, France}

\begin{document}
\maketitle

\begin{abstract}
In astronomy and microscopy, distortions in the wavefront affect the dynamic range of a high contrast imaging system. These aberrations are either imposed by a turbulent medium such as the atmosphere, by static or thermal aberrations in the optical path, or by imperfectly phased subapertures in a segmented mirror. Active and adaptive optics (AO), consisting of a wavefront sensor and a deformable mirror, are employed to address this problem. Nevertheless, the non-common-path between the wavefront sensor and the science camera leads to persistent quasi-static speckles that are difficult to calibrate and which impose a floor on the image contrast. In this paper we present the first experimental demonstration of a novel wavefront sensor requiring only a minor asymmetric obscuration of the pupil, using the science camera itself to detect high order wavefront errors from the speckle pattern produced. We apply this to correct errors imposed on a deformable microelectromechanical (MEMS) segmented mirror in a closed loop, restoring a high quality point spread function (PSF) and residual wavefront errors of order $\sim 10$~nm using 1600~nm light, from a starting point of $\sim 300$~nm in piston and $\sim 0.3$~mrad in tip-tilt. We recommend this as a method for measuring the non-common-path error in AO-equipped ground based telescopes, as well as as an approach to phasing difficult segmented mirrors such as on the \emph{James Webb Space Telescope} primary and as a future direction for extreme adaptive optics.
\end{abstract}


\section{Introduction}
\label{intro}

A major goal in present-day astronomy is the direct detection of planets and other faint companions to stars, a task which simultaneously requires high contrast and high resolution. This task is principally limited by the diffraction of light from the parent star, due to static aberrations in the telescope optics, quasi-static errors that vary with telescope pointing and environmental conditions, and the dynamic effects of atmospheric turbulence. 

Of these, by far the most severe difficulty is with atmospheric turbulence, or `seeing'. It has been apparent at least since Newton's times that atmospheric turbulence limits the resolution and sensitivity of astronomical observations. This problem can be substantially overcome by the use of adaptive optics (AO), first proposed by \citet{1953PASP...65..229B}. In this approach, a wavefront sensor is paired with a deformable mirror (DM) in a feedback loop to measure and correct for distortions in the phase of the incoming light \citep{2012ARA&A..50..305D}.  

The application of AO is not strictly limited to astronomy, however. In optical microscopy it is also the case that the specimen under study introduces aberrations into the optical path that limit the ability of a microscope to resolve detail deep below the surface of an otherwise transparent specimen \citep{Booth15122007}.

There is a third, related, problem in optics for which a wavefront sensor is necessary. The primary mirrors of large telescope are constructed out of segments. Current examples include the two 10-meter W.~M.~Keck Telescopes and the \emph{James Webb Space Telescope}. Future Extremely Large Telescopes (ELTS) will follow this design: the primary mirrors of the European ELT, the Thirty Meter Telescope (TMT) and the Giant Magellan Telescope (TMT) are all to be made of multiple elements. Discontinuities in the pupil and sharp segment edges however come with difficulties that are not well addressed by traditional wavefront sensors whose reconstructors assume continuity of the wavefront. It is therefore of prime importance to ensure that all segments are positioned correctly relative to one another without introducing significant wavefront errors. 

Hitherto the most common technology for wavefront sensing has been the Shack-Hartmann wavefront sensor. It uses an array of lenslets placed in a re-imaged pupil, such that the position offset of the spot produced by each lenslet encodes information about the local slope of the wavefront. This has been regularly applied to general wavefront sensing applications since \citet{1992ApOpt..31.6902L} and is now ubiquitous in astronomical AO \citep{2012ARA&A..50..305D}. However as noted by \citet{2005ApJ...629..592G}, the Shack-Hartmann technology is intrinsically vulnerable to photon noise and requires a large fraction of the incoming light be diverted for the sole purpose of wavefront sensing. This is a significant drawback in the regime of faint targets or high-contrast studies. 

Other technologies such as the pyramid wavefront sensor \citep{1996JMOp...43..289R}, curvature wavefront sensor \citep{1988ApOpt..27.1223R} and phase contrast method \citep{1934MNRAS..94..377Z} are possible. They however also suffer from similar drawbacks.

All aforementioned approaches suffer from the non-common-path (NCP) error problem, as aberrations occurring downstream from the wavefront sensor are not directly measured. These residual errors give rise to quasi-static speckles that impose a detection floor in high-contrast imaging, such as with the P1640 integral field spectrograph \citep{2008SPIE.7015E..32H,2011ApJ...729..132C,2011PASP..123...74H}. 

Solutions to this problem require some form of phase retrieval using the science camera itself. Possible implementations fall into two categories: the first relies on an active differential process. Diversity in the point spread function (PSF) is introduced by an active element to distinguish coherent, variable diffraction features from incoherent, fixed celestial objects. The other is a passive differential process. One standard approach representative of this category is angular differential imaging (ADI) \citep{2006ApJ...641..556M, 2007ApJ...660..770L}, where the pupil is allowed to rotate with respect to the sky (as in, for instance, an alt-azimuth mounted telescope). If this rotation occurs on a timescale shorter than that of the variation of the quasi-static speckles, it is possible to distinguish faint companions from speckles in the PSF, in post-processing.

To measure the NCP-error from the science detector, the standard option is the phase diversity technique. Multiple images taken in and out of focus, make it possible to determine uniquely the static aberrations across the entire wavefront \citep{1994ApOpt..33.6533K,2004OptL...29.2707C,2012OptL...37.4808S}. This NCP-error estimate can be used to offset the wavefront sensor zero-position for close-loop operation.

The DM itself can be used to introduce the known phase perturbation in order to calibrate or suppress speckles. The large number of degrees of freedom offered by a DM leads to variety of control algorithms: from random perturbations of the DM while monitoring a metric function of the PSF \citep{2012PASP..124..247R}; iterative speckle nulling loop compatible with coronagraphic mode \citep{2012PASP..124.1288M}; to a more complete determination of the wavefront with a finite number of DM modulations \citep{2012SPIE.8447E..21K}. A comparable method compatible with a coronagraphic imaging mode should rely on the electric field conjugation framework \citep{2007SPIE.6691E...7G,2010lyot.confE..63T,2011SPIE.8151E..32G}, to construct a complex transfer matrix for the real and imaginary parts of the electric field between pupil and focal planes, for direct speckle nulling within a finite region of the field of view.
These approaches all rely on sequential motions of the DM to achieve the required diversity, which is necessarily time-consuming and therefore a challenge for busy observing schedules.

The alternative to this is interferometric calibration, as is already done with sparse aperture masking with AO \citep{2006SPIE.6272E.103T}. At the cost of Fourier coverage and throughput, it is possible to extract closure phases which are self-calibrating with respect to phase errors in the pupil. Because they are measured from the science camera image, closure-phase are robust against residual aberrations, NCP-error otherwise. The idea of closure phases can be generalized to `kernel phases' for arbitrary pupils in the limit of a well-corrected wavefront \citep{2010ApJ...724..464M}. This approach has proven fruitful in detecting faint companions to brown dwarfs in the `super resolution' regime, beyond the formal diffraction limit, and with the calibration schemes proposed by \citet{2013MNRAS.433.1718I} can complement ADI and similar techniques at small angular separations. Kernel-phase uses a matrix pseudoinverse approach and therefore has formal features in common with electric field conjugation. The response matrix is however not empirically determined, and relies instead on a simple model of the pupil geometry, while assuming that amplitude errors are negligible. This has the advantage of not requiring calibration of the same extent or duration.

In this paper we present the first laboratory demonstration of the asymmetric pupil Fourier wavefront sensor proposed by \citet{2013PASP..125..422M}. This emerges naturally as a dual to the kernel phase image analysis method put forward by \citet{2010ApJ...724..464M}. By considering the problem of PSF formation from an interferometric perspective, it is possible to recover with post-processing a map of the wavefront giving rise to the PSF of any inversion-asymmetric pupil subject to weak aberrations.  This incurs a modest hardware cost of a minor asymmetric obscuration of the pupil. For segmented mirrors, this may be achieved by segment tilting or by a judicious arrangement of a mask or spiders. The advantage here is that this is in principle a single-step, requiring no phase diversity or modulation.

Using simulations, \citet{2013PASP..125..422M} shows that the method is particularly adapted to the determination (and therefore correction) of the NCP-error in an AO system. Here, we provide experimental results showing that it is equally suited to the phasing of a segmented mirror.

\section{Phase Transfer Algorithm}
\label{algorithm}

\subsection{Theory}
\label{theory}

Here we will summarize and explain the theory of the wavefront sensing approach proposed by \citet{2013PASP..125..422M}. In the following discussion it will be important to distinguish between three planes in the optical path: the pupil plane, defined at the telescope aperture; the image plane, as recorded (for instance) by a camera; and the Fourier plane, the Fourier transform of the image. The image of a point source (or equivalently the point spread function of the imaging system) is given by the squared Fourier transform of the electric field in the pupil, and therefore the complex visibilities in the Fourier plane are given by the field's autocorrelation. For arbitrary aberrations, it is not in general possible to uniquely map from information in the image plane only to the wavefront in the pupil plane, because the autocorrelation of an arbitrary function is not uniquely invertible. 

\begin{figure}[h!]
\center
\includegraphics[scale=0.4]{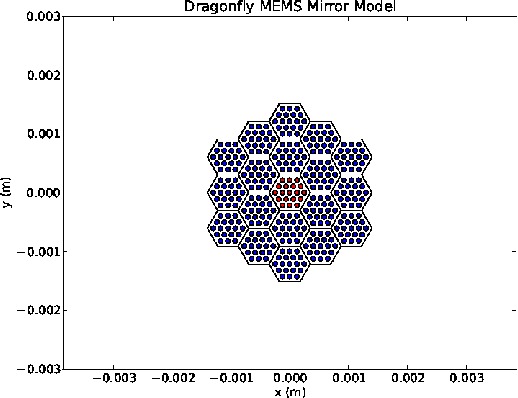}
\caption{A model of the \emph{Dragonfly} segmented mirror. Blue points represent points tilted on-axis; red points, an equivalent sampling to illustrate mirrors tilted away from the optical axis.}
\label{fullpupil}
\end{figure}

If the aberrations are small ($\bm{\varphi} \lesssim 1~\text{rad}$), it is however possible to treat this mapping as a linear operation.
Using a discrete model of the pupil it is then straightforward to compute the phase transfer matrix associated with this operator. In this paper, will consider models of the hexagonal segmented MEMS mirror used for the Dragonfly pupil remapping interferometer \citep{2009OExpr..17.1925K}, which is a scale model of the \emph{JWST} pupil. A filled pupil is shown in Figure~\ref{fullpupil}. 

\citet{2010ApJ...724..464M} showed that the singular value decomposition (SVD) of the transfer matrix is separable into mappings between pairs of orthonormal phase vectors in the pupil and Fourier planes, and a kernel space of vectors in the Fourier plane to which no small perturbations in the pupil plane can be mapped.  

In particular, we can write this expression as.

\begin{equation}
\mathbf{\Phi} = \mathbf{A}\cdot\bm{\varphi} + \mathbf{\Phi_0}
\end{equation}

for a transfer matrix $\mathbf{A}$, pupil phases $\bm{\varphi}$, observed Fourier phases $\mathbf{\Phi}$ and `true' source Fourier phases $\mathbf{\Phi_0}$. A simple method for obtaining the matrix $\mathbf{A}$ is described in \citet{2010ApJ...724..464M}. We then obtain the SVD of $\mathbf{A}$, 

\begin{equation}
\mathbf{A} = \mathbf{U}\cdot\mathbf{\Sigma}\cdot\mathbf{V^T}
\end{equation}

The left-annihilator of $\mathbf{A}$, $\mathbf{K}$, is then spanned by the columns of $\mathbf{U}$ corresponding to singular values of zero.  It is then possible to extract kernel phases such that: 

\begin{align}
\mathbf{K}\cdot \mathbf{\Phi} &= \mathbf{K}\cdot\mathbf{A}\cdot\mathbf{\varphi} + \mathbf{K}\cdot\mathbf{\Phi_0}\nonumber\\
\therefore\mathbf{K}\cdot \mathbf{\Phi} &= 0 + \mathbf{K}\cdot\mathbf{\Phi_0}.
\end{align}

By extracting these kernel phases $\mathbf{K}\cdot \mathbf{\Phi}$ from high-Strehl astronomical observations, such as from space telescopes or with assistance from extreme adaptive optics, it is possible to dramatically enhance the signal-to-noise of phase information contributed by real asymmetries in an astronomical source. Even in the case of a nominally diffraction-limited space telescope, low-level thermal and vibrational modes of the telescope optics nevertheless contribute to the degradation of the wavefront quality and therefore the PSF. Using kernel phase analysis, \citet{2013ApJ...767..110P} were able to obtain high resolution, high-contrast information at and beyond the formal diffraction limit of the \emph{Hubble Space Telescope}. 

The remaining phase vectors in the SVD can be used to construct a Moore-Penrose pseudoinverse of the phase transfer matrix, so that small aberrations in the wavefront can be uniquely reconstructed from the Fourier plane \citep{2013PASP..125..422M}. In the following, we will briefly outline the inversion method. 

It is crucial in this case that the pupil itself not be centrosymmetric: that is, that under the transformation $(x,y) \rightarrow (-x,-y)$ it does not map onto itself. This is because otherwise it is not possible to distinguish between pupil modes of odd and even parity under inversion.

With a symmetric pupil, only odd modes of the pupil appear in the row space and can be sensed when constructing the pseudoinverse. Introducing a pupil plane asymmetry, even of a very small character, breaks this degeneracy, and the pseudoinverse of this transfer matrix will accordingly map to the full space of pupil modes.

\begin{figure*}
  \centering
  \subfigure{\includegraphics[scale=0.4]{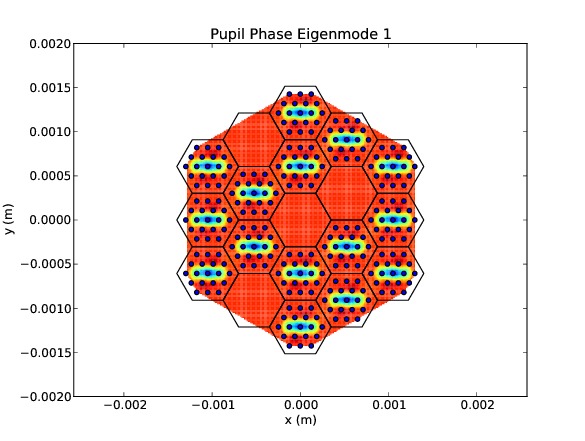}}\quad
  \subfigure{\includegraphics[scale=0.4]{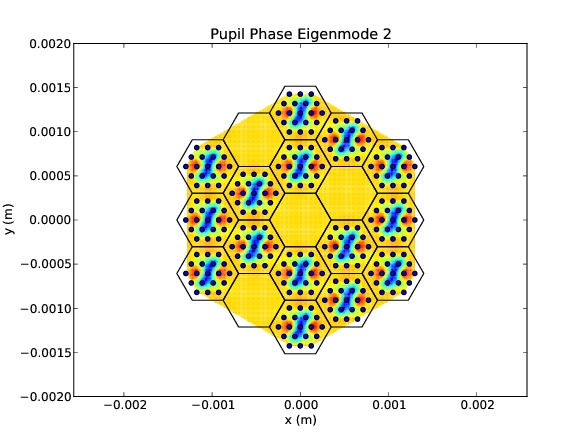}}\quad
  \caption{Examples of pupil mode patterns calculated for the first non-redundant mask geometry as shown in Figure~\ref{MEMSlayout}, interpolated onto the convex hull of the pupil sample set. Colour scale is arbitrary.}
  \label{modes}
\end{figure*}

The column vectors in $\mathbf{U}$ provide a set of orthonormal modes for describing phases in the pupil. The transfer matrix $\mathbf{A}$ maps these one-to-one onto a corresponding orthonormal basis of row vectors in $\mathbf{V}$, normal modes for the Fourier plane. These can be thought of as being similar to a symmetry-adapted set of Zernike-like modes for an arbitrary pupil. The Moore-Penrose pseudoinverse $\mathbf{A^+}$ maps the vectors back in the opposite direction:

\begin{equation}
\mathbf{A^+} = \mathbf{V}\cdot\mathbf{\Sigma^+_k}\cdot\mathbf{U^T}
\end{equation}

where $\mathbf{\Sigma^+_k}$ denotes the diagonal matrix whose entries are the reciprocals of the first $k$ diagonal entries of $\mathbf{\Sigma}$. 
Examples of these modes are displayed in Figure~\ref{modes}.

\subsection{Software Implementation}
\label{code}

In this experiment we have used a Python implementation of the asymmetric pupil Fourier wavefront sensing algorithm, sharing several common features with the \texttt{pysco} `PYthon Self-Calibrating Observables' kernel-phase analysis code. We calculate the pseudoinverse from the SVD of the transfer matrix $\mathbf{A}$, including only the first 200 singular values and corresponding modes out of a total number of $\sim 1000$. This cutoff at 200 modes was chosen in order to help smooth the resultant wavefront estimate, as the mirror phasing problem requires predominantly low-order information 

While $\mathbf{A}$ is in general sparse, we take care not to na\"{i}vely apply sparse matrix algebra packages, and instead treat the full matrix in all operations. We do this in order to preserve small components of the normal modes, which we found were truncated in the sparse treatment in such a way as to prevent the convergence of the algorithm. 

The algorithm proceeds as follows. The image is loaded, bias-subtracted, re-centred, the Fourier transform taken, and sampled at the baselines generated by the pupil model. We then operate on these $uv$ phases with the transfer matrix pseudoinverse, to obtain a pupil wavefront map. Next, we fit and subtract an overall tip-tilt from this map, to account for imperfections in the recentring. Finally, we obtain the piston, tip and tilt on each segment from the mean, $x$ gradient and $y$ gradient of the wavefront across the sampling points.

In all practical cases, these operations are very fast. For a pupil sampling as dense as in Figure~\ref{fullpupil}, this takes of order several seconds on a laptop with Intel(R) Core(TM) i7-2760QM CPU running at 2.40GHz. Likewise, the entire process of extracting a wavefront from a fresh image frame can take as little as two seconds for such a sampling.

We have not attempted to optimize the computational speed of this algorithm, which is implemented in pure Python. As the density of a pupil sampling increases, the number of $uv$ baselines grows rapidly, and although required only once, the SVD may be very slow, or fail to converge. Precalculation of these matrices is therefore essential.

After this, processing time for a typical single frame in this experiment is set by the Fast Fourier Transform (FFT) of a 1024 x 1024 image, sampling at the 1872 unique baseline points and multiplying this vector by a 1874 x 284 matrix. This took $\sim$~2~s on our operating laptop. The FFT step takes 76\% of the calculation time in this case, which we suggest could in AO applications be significantly improved with more efficient algorithms and hardware. A high order AO system such as PALM-3000 or SCExAO on a 5-10 m telescope needs to operate at  $\sim$ kHz rates \citep{2012ARA&A..50..305D} and it is not unreasonable to expect this to be achievable.

Future applications should rely on a library of pupil models pre-computed with high-performance methods. With such high-performance resources, this approach to wavefront sensing will be compatible for a very fast closed AO loop. For this experiment, such a link was maintained between computers controlling the camera and running analysis software using a file-sharing client. In future implementations, we recommend both operations be conducted on the same computer. 

\section{Apparatus}
\label{apparatus}

\begin{figure}
\center
\includegraphics[scale=0.17]{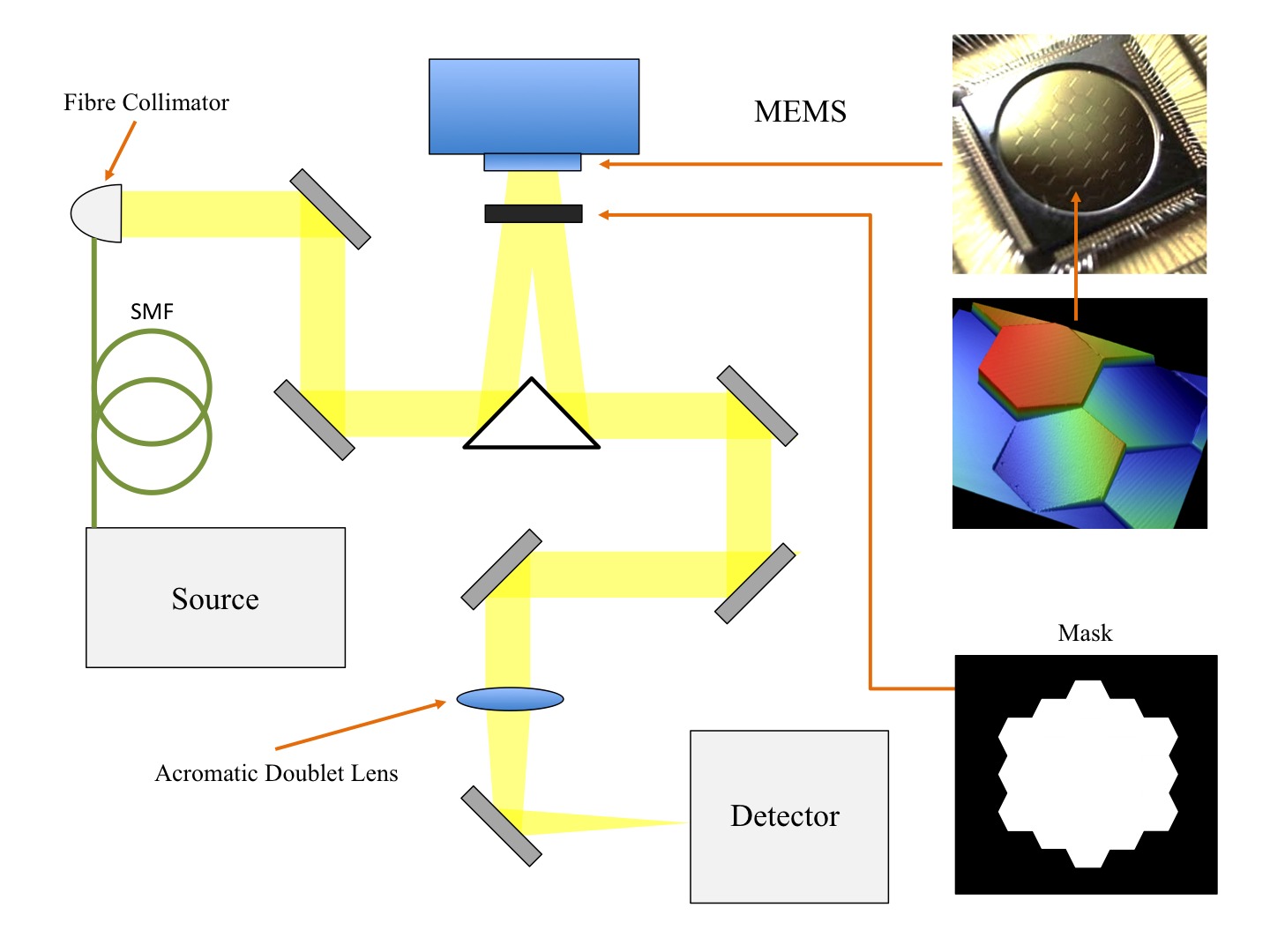}
\caption{Experimental layout used for testing the segment phasing algorithm. A single mode laser source was collimated onto the MEMS, passing through a mask to remove unwanted stray light from the periphery. Each segment was controlled in tip, tilt, and piston. The light was then focused onto an InGaAs NIR detector using an achromatic doublet lens forming the image. Inset images of MEMS from \citet{2011SPIE.7931E...7H}.}
\label{fig:setup}
\end{figure}

The experimental setup used to test the algorithm is shown in Figure \ref{fig:setup}. A Micro-Electro-Mechanical Segmented Mirror (MEMS) served as an analogue of a typical telescope segmented primary mirror. The MEMS consists of $37$ hexagonal, gold-plated mirrors arranged in a 4-ring hexagonal close-packed configuration, which can be adjusted electronically in piston, tip, and tilt, to a precision of a few nm (Iris AO PT-111 DM \citet{2011SPIE.7931E...7H}). Each hexagonal segment is $700~\mu m$ corner-to-corner (or $606.2~\mu m$ flat-to-flat), with the whole MEMS active area measuring $\sim 4.2~mm$ in diameter. The MEMS was illuminated by a narrowband laser source (Tunics Tunable C-band laser) at a wavelength of $1600$~nm, which was injected into a single-mode optical fibre (SMF-28), and collimated using a reflective fibre collimator (a $90 ^{\circ}$ off-axis parabolic mirror). The reflected light from the MEMS formed an image on an InGaAs detector array (Xenics Xeva 1.6-640 NIR Camera), focused using a $200$~mm focal length NIR achromatic doublet lens (Thorlabs AC254-200-C-ML). A number of silver coated mirrors with tip-tilt adjustment were used in the optical path to steer the beam, ensuring on-centre incidence for all key optical components.   

\begin{figure*}
\center
\includegraphics[scale=0.35]{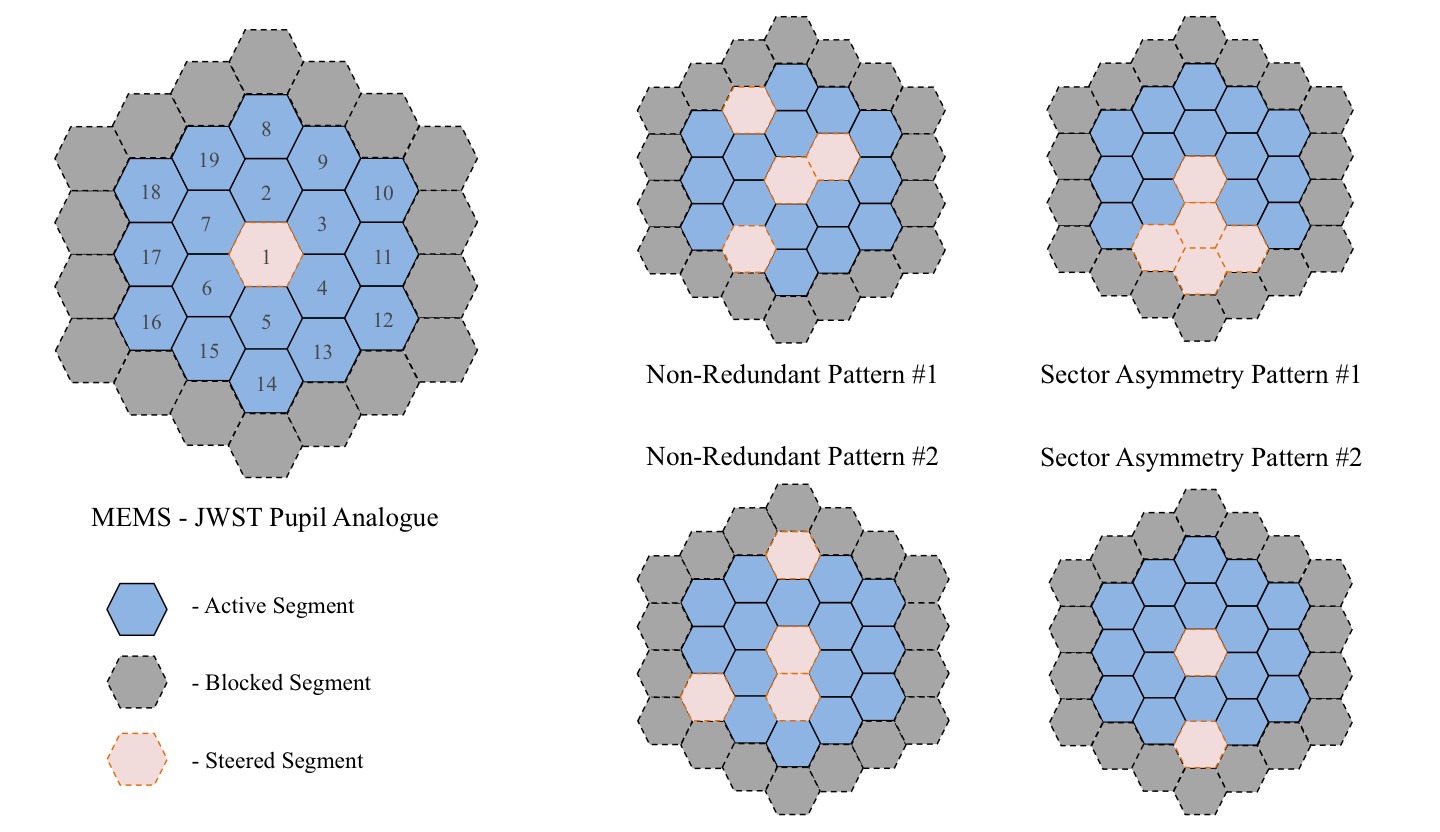}
\caption{Schematics of the various MEMS pupil patterns used to test the algorithm. For all tests, the outer ring of segments was blocked by a pupil mask (gray segments) and did not contribute light to the final image. The segments which were unwanted were steered away by tilting the segments to their maximum travel such that they were not focused onto the NIR detector (pink segments). The remaining segments (blue) were used to form the image. This technique enables the creation of arbitrary segment patterns in the pupil to assess the impact of symmetry-breaking required for convergence, as well as approximating spider layouts in the pupil. }
\label{MEMSlayout}
\end{figure*}

As mentioned above, the MEMS acted as a scale model for the $JWST$ pupil. However, because the MEMS has an additional outer ring of segments when compared to the $JWST$ primary, a mask was placed at a distance of $<1$~mm in front of the MEMS which blocked out the unwanted segments. To further match the $JWST$ pupil, a tilt was applied to the central segment such that it did not contribute any light to the image plane, steering the beam away from the centre of the detector. The ability to tilt away unwanted MEMS segments allowed for an arbitrary layout for various pupil-shapes to be created for testing the algorithm's performance. Examples can be seen in Figure \ref{MEMSlayout}.

Our apparatus differs from $JWST$'s geometry in another important respect: while we have tilted away the central segment which would be blocked by the secondary mirror of the real telescope, we have not attempted to include the spiders which support this secondary mirror. One should note that with three spiders, the $JWST$ pupil is already asymmetric, and it will be useful in future to establish whether this in itself provides enough asymmetry for our wavefront sensing scheme. This remains beyond the scope of the present work. This may raise the question as to whether the presence of spiders may actually harm the performance of this method. Since the spiders contribute to the pupil structure only at very high spatial frequencies, we do not expect them to contribute detrimentally to wavefront sensing at the low spatial frequencies required in the mirror phasing problem.

\section{Results}
\label{sec:res}

In the following Subsections~\ref{nr}-\ref{noasym} we will discuss the results of the different wavefront sensing experiments performed: in Section~\ref{tests} we describe initial tests establishing the effectiveness of the technique; in Section~\ref{nr}, we phase an entire mirror tilting away a scalene array of three segments each time; in Section~\ref{wedgeandsingle}, removing a whole quadrant of the full mirror, then with only a single mirror missing at the edge; and in Section~\ref{noasym}, without altering the mirror structure and simulating the asymmetry in the pupil sampling. In all cases with real pupil asymmetry we recover a high-quality PSF from an initially-heavily-degraded map; and in the case with asymmetric sampling but a full pupil, we fail to achieve any significant restoration of the PSF. 

Unfortunately, the NIR camera used in these experiments had a far lower dynamic range, and far higher noise floor, than that of state-of-the-art NIR detectors used on modern telescopes. As such, visualising the faint airy rings without saturating the central spot was not possible. This is further complicated by high detector non-linearity at both the low, and high, ADU ranges. While we have corrected for these, this process will have induced small extra errors in our measurement.

\subsection{Initial Tests}
\label{tests}

\begin{figure}
\center
\includegraphics[scale=0.17]{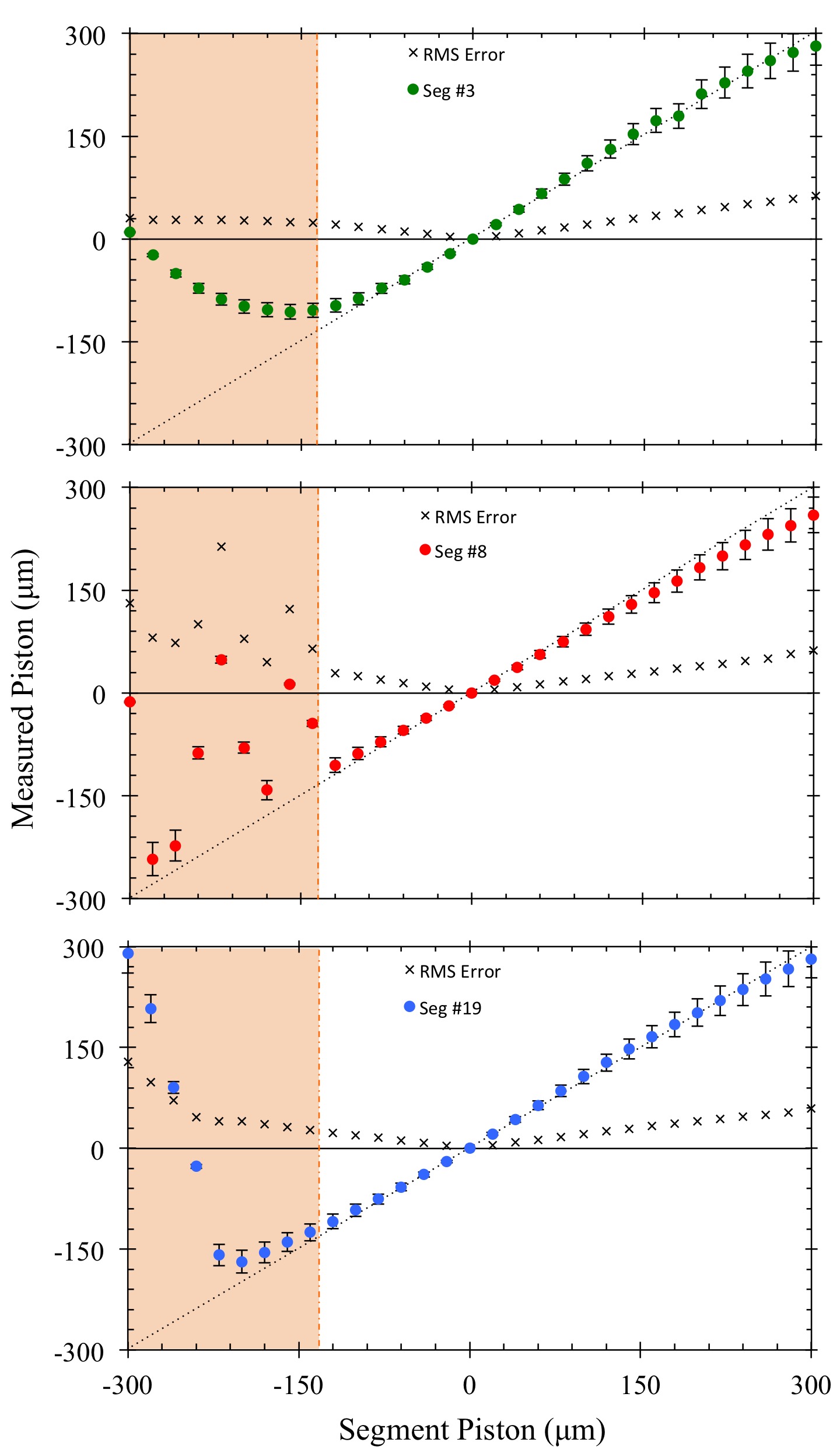}
\caption{Measured segment pistons using the wavefront sensor for segment \#3 (top), \#8 (middle), \& \#19 (bottom). The RMS piston values of the remaining unpistoned segments are also shown. The highlighted orange section of the graph shows the region which is outside the 1~rad threshold (dominated by the cubic term and no longer linear), where the reconstruction has failed.}
\label{fig:pistongraph}
\end{figure}

In order to establish the effectiveness of the wavefront sensor, we initially tested its accuracy using the simplest possible cases. As mentioned in Section \ref{theory}, the sensor only functions if the pupil symmetry is broken in some manner. Hence, for our initial experiments we used the \textit{Sector Asymmetry Pattern $\# 1$} (see Fig \ref{MEMSlayout}.) by tilting away a scalene triangle of segments, providing the greatest possible pupil asymmetry.

With the pupil pattern established, we set all remaining MEMS segments to their mechanical zero positions, and reconstructed a wavefront from an exposure here. We then took this as our arbitrary zero wavefront for all segments. It is important to note that while the MEMS is able to zero the pistons of the segments to an accuracy of 10~nm, there are further wavefront errors and global tilts induced by downstream optics, for instance mirrors, lens, detector window, so that this zero-point wavefront map is not perfectly flat. However, because in the experiments presented in this subsection we measure a relative piston compared to an arbitrary starting position, and not an absolute piston, this differential technique is adequate to demonstrate the wavefront sensor's applicability.

MEMS segments 3, 8, \& 19 (see Fig.~\ref{MEMSlayout}.) were independently pistoned from $-300$~nm to $+300$~nm in $20$~nm increments. At each step an image was taken of the resulting PSF and the wavefront reconstructed by our algorithm. After the phase-map was normalized to the zero reference map, piston of all active segments were extracted. An important subtlety is that the measured wavefront piston is actually \emph{twice} the MEMS segment piston, because the light reflecting off the segment acquires an optical path length difference in both directions. 

Figure~\ref{fig:pistongraph} shows the accuracy of the wavefront sensor's piston measurement for the three segments that were pistoned. As can be seen in Fig.~\ref{fig:pistongraph}, the reconstructed piston on the correct segment tracks the input values very closely (within the measurement error of $10\%$) between $-135$~nm \& $+300$~nm of piston, outside of which ($<-135$~nm) it varies more or less wildly. This is in accordance with our expectation that for input errors larger than a threshold value of 1~rad the reconstruction should fail, as the cubic term dominates over the linear term. We note that this in practice occurred for relatively small negative values, but maintained its accuracy for positive values for as far as we tested; this may be owing owing to phase offsets from the zero-phase reference level, which was biased such that the `error centre' took a positive value. Also shown in Fig.~\ref{fig:pistongraph} is the mean residual measured pistons from the remaining unpistoned segments.  

We note that the wavefront recorded for the other segments varies in proportion to the pistoned segment. This gives rise to an RMS wavefront error across the pupil that grows monotonically with any local error. We conjecture that the reason for this is the choice of basis. The phase is projected onto a truncated basis of modes supported on a limited, discrete set of points. Accordingly a large phase error near an anti-node of one of these modes is liable to bias the reconstruction across the whole pupil. Because this bias is monotonic with the input, when a negative feedback loop is applied, as in the following experiments, it will in general be ironed out as the iterations progress and will not significantly affect convergence of the algorithm.

\subsection{Non-Redundant Triplet Asymmetries}
\label{nr} 

As noted in Section~\ref{theory}, the singular value decomposition constructs two basis sets to span the pupil and Fourier planes. While the method only requires that the pupil possess no inversion symmetry, it is apparent from simple inspection of the generated pupil modes that they are by construction symmetry-adapted to any other symmetries present in the pupil, most notably lines of reflection symmetry.

It therefore seemed most promising for an initial test to phase a pupil with no symmetries at all. One choice is to remove first one scalene triangle of segments, and then bring these back in and remove a complementary triplet non-redundant with respect to the first. In this manner, all Fourier components and all mirror segments are sensed, and there are no spatial symmetries in the pupils used for sensing.

The remaining mirrors were set to their fiducial flat configurations. Then, random pistons and tip-tilts were drawn from normal distributions with variances of 300~nm and 0.3~mrad respectively, and applied to the MEMS controls. These errors were chosen such that at the 1600~nm wavelength the RMS error would be just beyond of the expected $\sim \lambda/{2\pi} = 250$~nm limit for the linear phase regime.  

\begin{figure*}
\center
\includegraphics[scale=0.3]{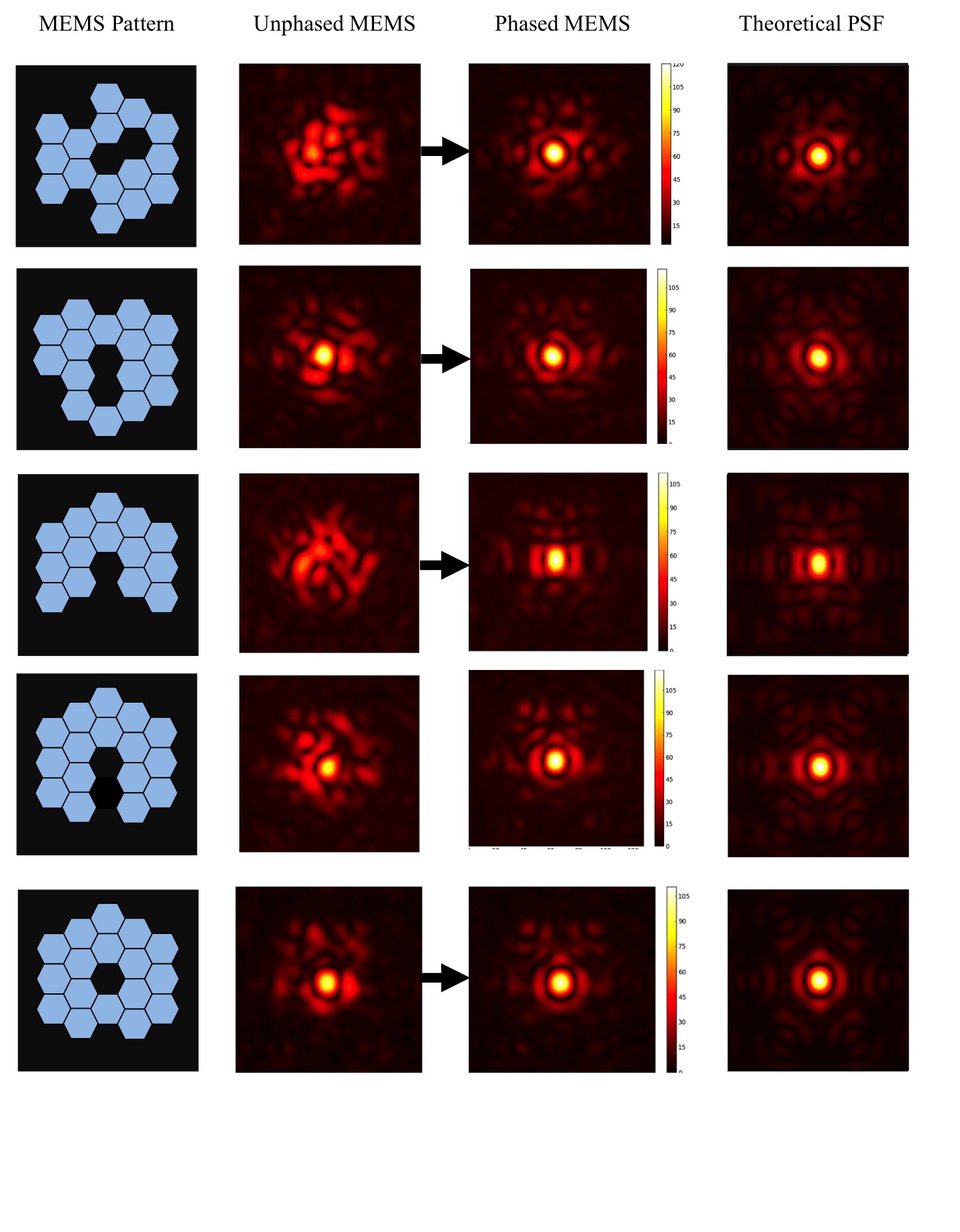}
\caption{Demonstration showing active phasing of the MEMS mirror using the wavefront sensor in a semi-closed loop. The left column identifies the active area of the MEMS, with specific segments tilted away to create different asymmetries (as described in more detail in Fig.~\ref{MEMSlayout}). The middle two columns show the uncorrected PSF, formed by randomly inducing pistons, tip, and tilts to the MEMS, and the final corrected PSF after running the phasing algorithm for 10 loops. The far right column shows the ideal theoretical PSF for the particular MEMS geometries. All images are fifth-root stretched such that the Airy pattern is clearly visible.}
\label{resultsgraphtest}
\end{figure*}

Starting the loop from this point, we ran the PSF restoration loop first for the first non-redundant pattern shown in Figure~\ref{MEMSlayout}, which removed segments 3, 15, \& 19. As shown in Figures~\ref{resultsgraphtest}, the PSF was quickly restored. 

In order to phase the entire mirror including the inactive segments, we tilted these back to their fiducial zero positions and tilted away three phased segments as shown in Pattern~2 of Figure~\ref{MEMSlayout}. We repeated the procedure here, and similarly obtained a high quality PSF.

By then restoring the settings of the segments removed in the first pattern, we were then able to phase the entire mirror. Unfortunately, the final phased mirror configuration led to saturation of the central pixels, which was noticed only in subsequent analysis. Nevertheless, it is visually apparent from inspection of the PSF that the diffraction pattern agrees extremely well with that calculated for the diffraction limit of a flat pupil.

\subsection{Wedge and Single Mirror Asymmetries}
\label{wedgeandsingle}

We repeated the previous experiment for two other asymmetries: first, with the wedge asymmetry as shown in Figure~\ref{MEMSlayout}, and second, with the single-mirror asymmetry. In both cases the algorithm converged to a high-quality PSF as with the non-redundant pattern. 

During the first attempt with the wedge asymmetry, we noted that one mirror's piston position as recorded by the controller was approximately 800~nm away from the others, that is, half the wavelength of the 1600~nm laser used as the light source. This suggested that while the wavefront appeared flat under this monochromatic source, this was because the Fourier wavefront sensing approach is insensitive to a $2\pi$ wrapping in phase. We therefore manually adjusted this mirror 800~nm down, near to the settings for the other mirrors, and found that the algorithm quickly converged again to a diffraction-limited PSF. 

The results of both experiments are displayed in Figure~\ref{resultsgraphtest}.

In this case, for both pupil configurations no such saturation occurred, and we were able to estimate the final Strehl ratio. In order to avoid a bias from residual scattered light, we did not directly calculate the overlap integral with a diffraction-limited PSF, but rather calculated radial encircled energy profiles for measured and theoretical images. These were calculated for the diffraction limit, and for a 99.0\% Strehl ratio PSF simulated for the manufacturer's quoted position accuracy limit for the MEMS, with segment positions drawn from a uniform distribution between $\pm$ 10~nm in piston and $\pm$ 0.05 mrad in tip-tilt. Simulated PSFs with 97\% Strehl ratio were a poor fit to the experimental profiles. We therefore conclude that the final Strehl $S$ was in the range $0.97<S<0.99$. Nevertheless, for all images there were departures at large radial distances, which we suggest are due to contamination with stray light. From these calculations we argue that in the inner regions of the PSF we attain performance limited primarily by the tolerance of the MEMS positioning, and closely approach the diffraction limit.

\subsection{No Asymmetry}
\label{noasym} 

In Section~\ref{theory}, it was stated that the matrix pseudoinverse does not span the whole range of symmetric and antisymmetric modes of the pupil unless the pupil model itself is asymmetric. From this it is not immediately apparent that an asymmetric sampling of a symmetric pupil would not permit the sensing of all modes. 

In order to test this, we tilted all mirrors on-axis apart from the centre panel, and deleted one point from the pupil sampling in order to make it asymmetric and obtain a matrix with modes spanning all the required modes. After randomizing the on-axis mirror settings as previously, we attempted to phase the mirror as before. After seventeen iterations there was no sign of improvement at all in the PSF, and we concluded that the method was ineffective in this case. 

This is to be expected on theoretical grounds, as for an even mode of an even pupil, the phase of the autocorrelation vanishes and we expect no signal in the Fourier plane, irrespective of the pupil model adopted in computation.

\section{Conclusions}
\label{conclusions}

This paper provides experimental evidence that the focal-plane wavefront sensing technique proposed by \citet{2013PASP..125..422M} is sound: even for a segmented aperture, it is possible to directly sense aberrations from the analysis of a single aberrated PSF, if one introduces some asymmetry in the pupil.
From the above results, it is clear that largely independent of the degree or structure of asymmetry, mirror phasing was rapid and effective. It is not possible based only on these results to suggest an optimal asymmetric mask strategy; on the other hand, it is clear that the algorithm is very forgiving with respect to the pupil geometry. It may accordingly be a fruitful direction for future simulations to optimize the pupil mask geometry and any effect that telescope spiders may have. 

This new wavefront sensing method is immediately applicable to several existing and near-future systems where a high-quality wavefront is degraded by quasi-static wavefront errors arising from small optical aberrations. As this experiment has shown, the problem of phasing a segmented mirror such as the upcoming James Webb Space Telescope is easily solved by this method. While the initial optical path error following unfolding is likely to be of order $\sim$ tens of $\mu$m, and other methods will be required to achieve coarse phasing, our approach is suitable as a `tweeter' on top of this by which the mirror shape can be fine-tuned and maintained. In this way, it will be possible to have an equivalent of `active optics' to maintain a uniformly high quality wavefront by adjusting the $JWST$ segments, requiring only that it briefly observe a bright point source. 

A second example in which this method can be applied is in correcting static NCP-error on ground-based telescopes with AO. This NCP-error is in many high contrast imaging applications the limiting error source (e.g. in \citet{2011ApJ...729..132C}) and a variety of solutions have been proposed \citep{2012SPIE.8447E..14T,2013A&A...555A..94N}. Few of these, with the exception of \citet{2012PASP..124..247R}, avoid the necessity of introducing substantial additional hardware into the beam path, with the attendant costs of time, expense and residual optical errors. \citet{2012PASP..124..247R} present a novel and effective algorithm for attaining a particular desired PSF, but the method demonstrated here has the advantage of producing \emph{a flat wavefront} quickly, for an arbitrary pupil, without model-dependent distortions and suitable for general observing. 

In applications where the readout speed of a camera can be made much faster than the atmospheric coherence time $t_0$, the technique demonstrated here would appear an ideal method of high-order wavefront sensing. This will be the case in any situation in which speckle interferometry might ordinarily be done, and is accordingly restricted only to bright targets. 

\section*{Acknowledgements}
\label{acknowledgements}
This research has made use of NASA's Astrophysics Data System.

This research was supported by the Australian Research Council Centre of Excellence for Ultrahigh bandwidth Devices for Optical Systems (project number CE110001018).

Part of this work was performed at the OptoFab node of the Australian National Fabrication Facility (ANFF).

\bibliographystyle{aa}
\bibliography{ms}

\end{document}